\documentclass[12pt]{article}
\include{epsf}

\newcommand{\eqn}[1]{(\ref{#1})}
\def\appendix#1{\addtocounter{section}{1}\setcounter{equation}{0}
\renewcommand{\thesection}{\Alph{section}}
\section*{
\thesection\protect\indent \parbox[t]{11.715cm} {#1}}
\addcontentsline{toc}{section}{Appendix\thesection\ \ \ #1} }
\newcommand{\real}{{\bb R}} 

\def\bra#1{\left\langle #1\right|}
\def\ket#1{\left| #1\right\rangle}
\def\hs#1#2{\left\langle #1\right|\left. #2\right\rangle}

\font\mybb=msbm10 at 12pt
\def\bb#1{\hbox{\mybb#1}}

\newcommand{\Tr}[1]{\:{\rm Tr}\,#1}

\def\be{\begin{equation}}
\def\ee{\end{equation}}
\def\bea{\begin{eqnarray}}
\def\eea{\end{eqnarray}}

\newcommand{\del}{\partial}

\newcommand{\dm}[2]{\ket{#1}\bra{#2}}
\newcommand{\pt}{\varphi^{\rm Tay}}

\begin{document}
\begin{flushright}

\baselineskip=12pt
DSF--34--2003\\
hep-th/0309128\\
\hfill{}\\
\end{flushright}

\begin{center}
\baselineskip=24pt

{\large{\bf From the fuzzy disc to edge currents \\in Chern-Simons
Theory }}

\baselineskip=14pt

\vspace{1cm}

{{\bf  F.~Lizzi\footnote{Talk presented at ``Space-time and Fundamental
Interactions: Quantum Aspects'', conference in honour of A. P.
Balachandran's 65th birthday, Vietri sul Mare, Salerno Italy, May 2003; To
be published in  Mod. Phys. Lett.}, P.~Vitale and A.~Zampini}}
\\[6 mm]

{\it Dipartimento di Scienze Fisiche, Universit\`{a} di
Napoli {\sl Federico II}\\ and {\it INFN, Sezione di Napoli}\\
Monte S.~Angelo, Via Cintia, 80126 Napoli, Italy}\\{\tt
lizzi, vitale, zampini@na.infn.it}
\\[10mm]

\end{center}

\vskip 2cm

\begin{abstract}
We present a brief review of the fuzzy disc, the finite algebra
approximating functions on a disc, which we have introduced
earlier. We also present a comparison with recent papers of
Balachandran, Gupta and K\"urk\c{c}\"{u}o\v{g}lu, and of Pinzul
and Stern, aimed at the discussion of edge states of a
Chern-Simons theory.

\end{abstract}


\section{Introduction}

To be working at almost the same idea of another group of people
is a fairly common occurrence. But the fact that the organizer of
a conference to celebrate the sixthyfifth birthday of a colleague,
and the person to be so honoured, discover this fact at the
conference qualifies it as an unusual fact. Also unusual is the near
\emph{triple coincidence} of another talk presenting a similar
model, somehow \emph{dual} to the other. This is what happened at
the conference whose proceedings are collected in this volume.
Since our talk was based on ref.~\cite{fuzzydisc} which was
posted shortly after the conference, at the same time as the work
of Balachandran, Gupta and
K\"urk\c{c}\"{u}o\v{g}lu\cite{BalGuptaKurkcuoglu}, rather than
writing a faithful account of the talk, we present a comparison of
the common elements of these two papers, and also discuss the
similarities with the work of Pinzul and Stern\cite{PinzulStern}
which was also presented at the conference.

We will discuss \emph{fuzzy} approximations to spaces with
boundaries. The aim is the same as studies on the lattice: the
substitution of a continuous theory with a model with a finite
number of degrees of freedom, a matrix model. In the limit in
which matrices become large the approximation improves. The main
advantage of the fuzzy approximation is that the basic symmetries
of the original space are preserved. The archetype of these is the
fuzzy sphere\cite{Madorefuzzysphere}. Along the same lines other
fuzzy spaces have been built
\cite{GrosseStrohamer,Ramgoolam,thefuzz}. The aim of the talk was
to discuss the fuzzy approximation to the disc, which is presented
in more detail in\cite{fuzzydisc}. A fuzzy disc (presented in
section~\ref{se:disc}) is a subalgebra of the noncommutative
plane, which we present is section~\ref{se:plane}. We also show
that a slight modification of the construction of the fuzzy disc
can lead to a strip, as noted in
ref.~\cite{BalGuptaKurkcuoglu}. We also briefly discuss the
model\cite{PinzulStern} where, instead of a disc, a plane with a
point missing at the origin is constructed. In section~\ref{se:CS}
we introduce derivations on the fuzzy disc, which enable the
construction of Chern-Simons actions and show the presence of the
edge states.

\section{The Noncommutative Plane \label{se:plane}}
\setcounter{equation}{0}

The principle behind all fuzzy spaces is to consider a
noncommutative geometry, that is a  noncommutative algebra
generalizing the algebra of functions on a topological space, and
then a finite dimensional representation, or a finite dimensional
subalgebra, of this noncommutative algebra.

For the construction at hand consider  functions on $\real^2$,
with coordinates $x$ and $y$ or $z=z+iy$ and $\bar z =x-iy$. Then
quantize the plane associating operators to functions so to have $
[\hat x,\hat y]=i\frac\theta 2$. The quantized versions of $z$ and
$\bar z$ are the usual annihilation and creation operators, with
\be
[a,a^\dagger]=\theta . \label{aadcomm}
\ee
The parameter $\theta$ has  the dimension of a square length. It
does not have necessarily a physical meaning, like the distance
between sites in a lattice approximation.

The particular quantization we choose is a map $\Omega_\theta$
which to the function $\varphi(z,\bar z)$ associates the operator
$\hat \varphi$ as follows. Consider the Taylor expansion:
\be
\varphi(\bar z,z)=\pt_{mn}\bar z^m z^n  \ , \label{taylorphi}
\ee
to this function associate the operator
\be
\Omega_\theta(\varphi):=\hat\varphi=\pt_{mn}{a^\dagger}^m a^n  \ .
\label{taylorexp}
\ee
The inverse map is constructed defining the \emph{coherent} states
$ a\ket{z}=z\ket{z}$, then
\be
\Omega^{-1}_\theta(\hat\varphi)=\varphi(\bar z,z)=\bra{z}\hat
\varphi\ket{z}  \ . \label{Omegam1}
\ee
There is another useful basis on which it is possible to represent
the operators. Consider the number operator
\be
{\rm N}=a^\dagger a  \ , \label{defnumb}
\ee
and its eigenvectors which we indicate\footnote{Since coherent
states corresponding to the natural numers never appear we hope
there will be no confusion between coherent states and
eigenvectors of ${\rm N}$.} by $\ket{n}$: $ {\rm
N}\ket{n}=n\theta\ket{n} $. We can then express the operators with
a density matrix notation:
\be
\hat\varphi=\sum_{m,n=0}^\infty\varphi_{mn}\dm{m}{n}  \ .
\label{dmexp}
\ee
The elements of the density matrix basis have a very simple
multiplication rule:
\be
\ket{m}\hs{n}{p}\bra{q}=\delta_{np}\dm{m}{q}  \ . \label{densmult}
\ee
The connection between the expansions~\eqn{taylorexp}
and~\eqn{dmexp} is given by:
\be
a=\sum_{n=0}^\infty \sqrt{(n+1)\,\theta}\dm{n}{n+1}\ \ ; \ \
a^\dagger=\sum_{n=0}^\infty \sqrt{(n+1)\,\theta}\dm{n+1}{n} \ .
\ee
Applying \eqn{Omegam1} to the operator $\hat\varphi$ in the number
basis we obtain for the function $\varphi$ a new expression, in
terms of new coefficients:
\be
\varphi(\bar
z,z)=e^{-\frac{|z|^2}{\theta}}\sum_{m,n=0}^\infty\varphi_{mn}
\frac{\bar z^m z^n}{\sqrt{n!m!\theta^{m+n}}}  \ . \label{newtay}
\ee
The maps $\Omega_\theta$ and $\Omega_\theta^{-1}$ yield a procedure of going
back and forth from functions to operators. Moreover,  the product
of operators being noncommutative,  a noncommutative $*$ product
between functions is implicitly defined as
\be
\left(\varphi*\varphi'\right)(\bar z,
z)=\Omega_\theta^{-1}\left(\Omega_\theta(\varphi)\, \Omega_\theta(\varphi')
\right)  \ .
\label{stardiff}
\ee
This product (which is a variation of the Moyal-Gr\"{o}newold
product) was first introduced by Voros \cite{Voros}. We indicate
the algebra of functions on the plane with this product as ${\cal
A}_\theta$. In the density matrix basis, because
of~\eqn{densmult}, the product~\eqn{stardiff} simplifies to an
infinite row by column matrix multiplication:
\be
\left(\varphi*\varphi'\right)_{mn}=\sum_{k=0}^\infty
\varphi_{mk}\varphi'_{kn} \ .
\ee
When $\theta\to0$,  the $*$ product goes to the ordinary
commutative product. It is easy to see that also $\int d^2z\,
\varphi(\bar z,z)={\pi\theta}
\Tr\Phi={\pi\theta}\sum_{n=0}^\infty\varphi_{nn} $, where we have
introduced the matrix $\Phi$ with components $\varphi_{nm}$.

\section{The fuzzy Disc \label{se:disc}}

\begin{figure}[th]
\epsfxsize=2.5 in
\bigskip
\centerline{\epsffile{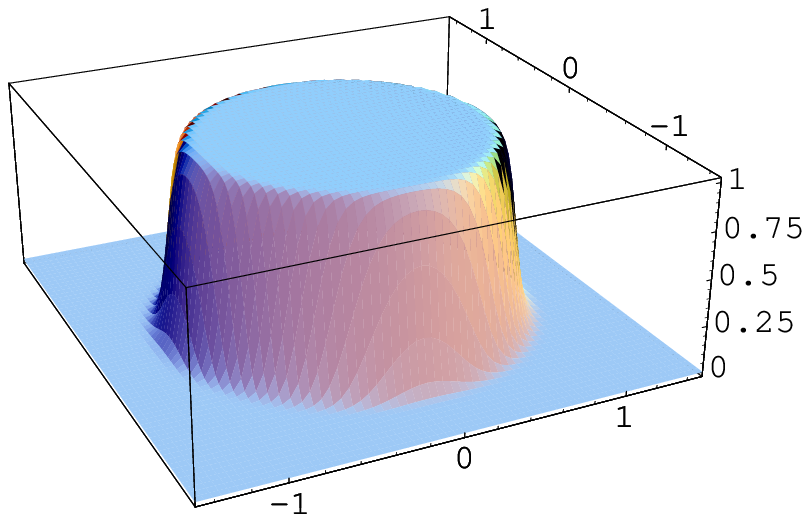}} \caption{\baselineskip=12pt
{\it The function $P^N_\theta$ for $N=10^2$.}}
\bigskip
\label{disc3d}
\end{figure}

\begin{figure}[th]
\epsfxsize=2.3 in \centerline{\epsfxsize=2.2
in\epsffile{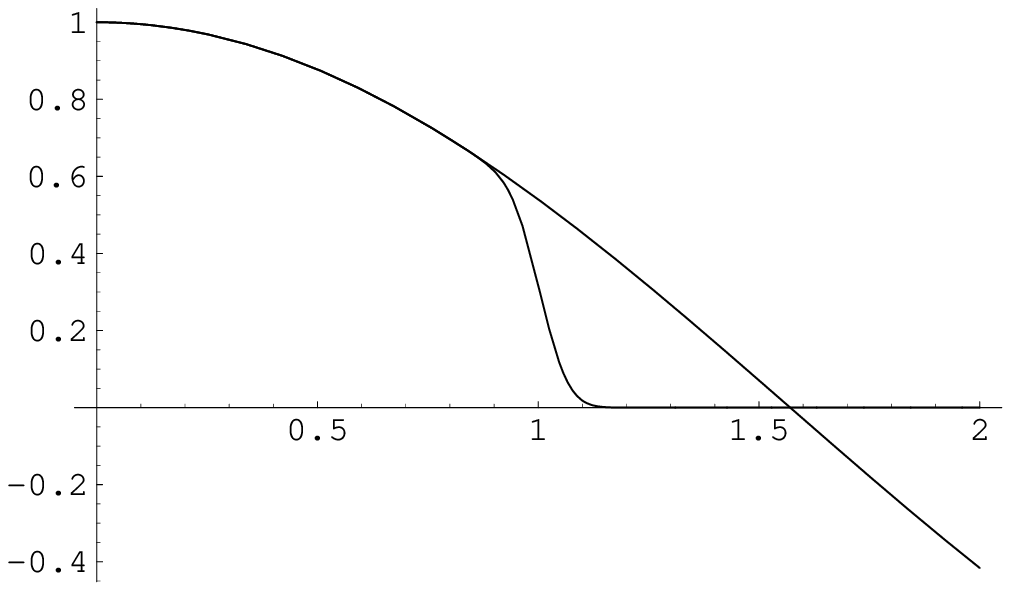}\epsfxsize=2.2
in\epsffile{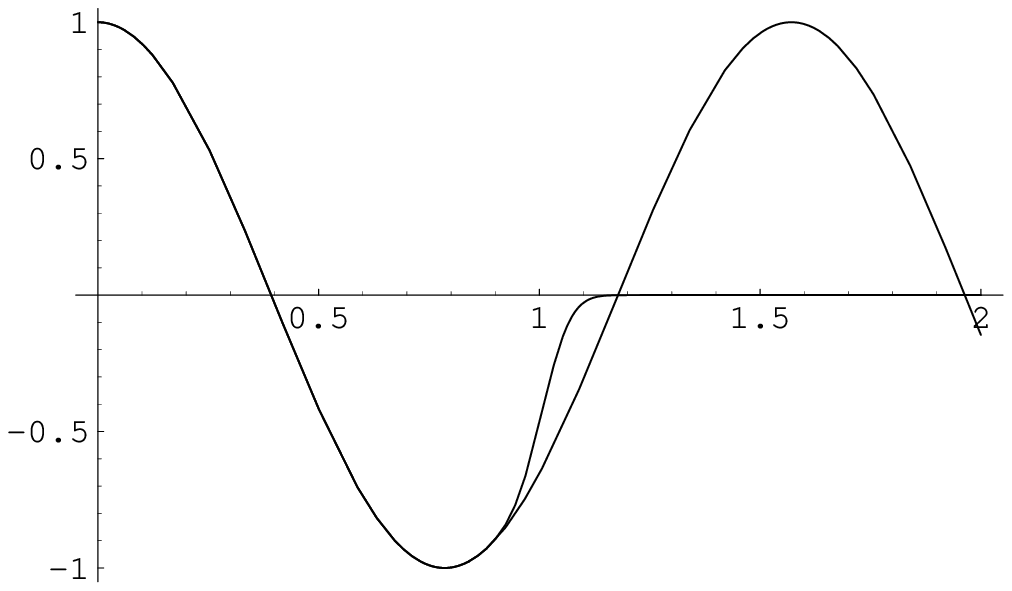}\epsfxsize=2.2 in\epsffile{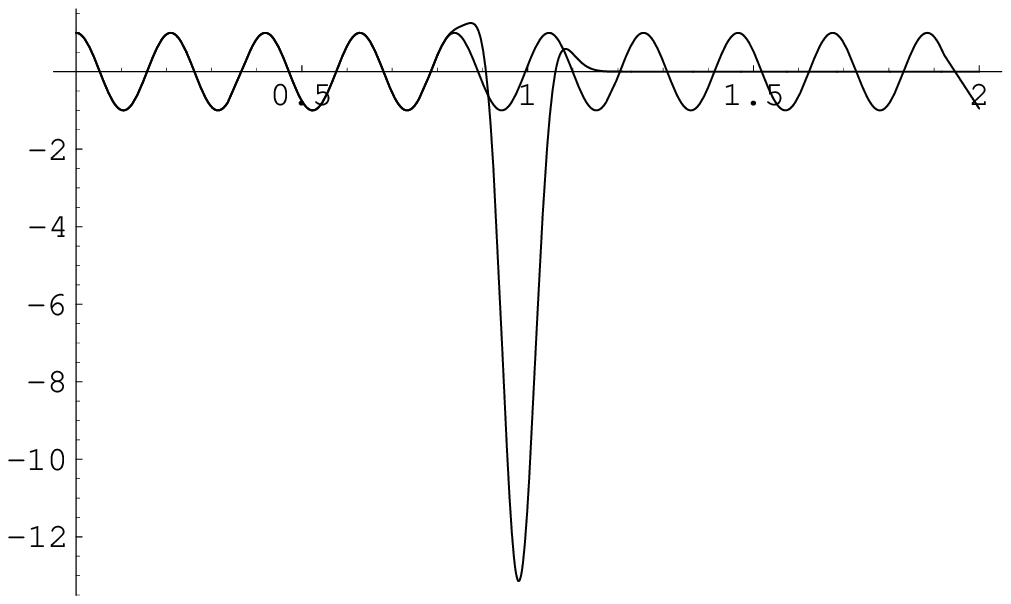}}
\caption{\baselineskip=12pt {\it Profile of the cylindrically
function symmetric function $P^N_\theta*\cos(\alpha r)*P^N_\theta$
for the choice, $N=10^2$ and for the values $\alpha=R, 4R, 30R$
compared with the unprojected function. Both functions are
plotted, although inside the disc they are often
indistinguishable.}}
\bigskip
\label{cosine}
\end{figure}

\setcounter{equation}{0} We now define  subalgebras (with respect
to the $*$ product) of \emph{finite} $N{\times} N$ matrices. They
are the functions whose expansion~\eqn{newtay} terminates when
either $n$ or $m$ is larger than a given integer $N$. They can be
obtained easily from the full algebra of functions via a
projection:
\be
P^N_\theta=\sum_{n=0}^N \hs{z}{n}\hs{n}{z}=\sum_{n=0}^N\frac{
r^{2n}}{n!\theta^n}e^{-\frac{r^2}{\theta}}  \ ,
\ee
where $z=r e^{i\phi}$. In the limit $N\to \infty$ and $\theta\to
0$ with
\be
R^2\equiv N\theta
\ee
fixed, the sum converges to $1$ if $r<R$, and converges to 0
otherwise. It has cylindrical symmetry. For $N$ finite the
function vanishes exponentially for $r$ larger than $R$ (see
figure~\ref{disc3d}) and approximates well the characteristic
function of the disc.
The function $P^N_\theta$ is a projector of the algebra of
functions on the plane with the $*$ product:
\be
P^N_\theta*P^N_\theta=P^N_\theta  \ ,
\ee
and the subalgebra ${\cal A}^N_\theta$ is defined as
\be
{\cal A}_\theta^N=P^N_\theta*{\cal A}_\theta*P^N_\theta  \ .
\ee
Cutting at a finite $N$ the expansion provides an infrared cutoff.
The cutoff is fuzzy in the sense that functions in the subalgebra
are still defined outside the cutoff, but are exponentially
damped. Moreover if one tries to localize a function on a scale
smaller than $\sqrt{\theta}$ then there is the appearance of large
values of the function on the boundary, a compact geometry version
of the ultraviolet--infrared mixing as  discussed in
ref.~\cite{fuzzydisc}. In figure~\ref{cosine} we show the
effect of projecting the function $\cos(\alpha r)$, that is we
plot $P^N_\theta*\cos(\alpha r)*P^N_\theta$ for various values of
$\alpha$. We see that for large values of $\alpha$ the projection
is just a cutoff, while for large $\alpha$ we see the presence of
a large peak on the boundary of the disc.
We call \emph{fuzzy disc} the space corresponding to the algebra
${\cal A}_\theta^N$, which is isomorphic to the algebra of
$N+1\times N+1$ matrices. It is important to note that what makes it a
disc is  the way to take the correlated limit of $\theta$ and $N$
keeping the dimensionful quantity $R$ fixed.

The procedure we just outlined can be in fact opportunely doctored
to obtain subalgebras which in the limit converge to functions
corresponding to portions of the plane with different shapes. This
can easily be obtained with just a redefinition of $z$ as
\be
z=\alpha_Nx+i\beta_Ny
\ee
with $\bar z$ defined accordingly, and $\alpha_N$ and $\beta_N$
appropriate functions of $N$. In general the projector has support
on an ellipse with semiaxes $\sqrt{N\theta}\alpha_N$ and
$\sqrt{N\theta}\beta_N$. This shape itself can change with $N$ and
give a different geometrical figure. The disc discussed earlier is
for $\alpha_N=\beta_N=1$ and $\theta=1/N$. A Noncommutative disc
would have $\alpha=\beta=1/\sqrt{N}$ with $\theta$ fixed, while a
noncommutative strip\cite{BalGuptaKurkcuoglu} is described by
$\alpha=1/\sqrt{N\theta}$ and $\beta=1$, which give an elongated
ellipse which in the limit becomes a strip. The choice
$\alpha_N=\beta_N=N$ gives the plane, and it must be noted that of
course those choices are not unique.

If we obtained the fuzzy disc subalgebra with the projector
$P^N_\theta$, we could have equally well considered the
projector $1-P^{N-1}_\theta$. In this case we have operators of the
kind:
\be
\hat\varphi=\sum_{m,n=N}^\infty\varphi_{mn}\dm{m}{n}  \ .
\ee
This describes the plane with missing a central portion, which
with an appropriate limit can be reduced to a point, and has been
introduced by Pinzul and Stern\cite{PinzulStern} in a different
way and well before refs.~\cite{fuzzydisc}
and~\cite{BalGuptaKurkcuoglu}. The algebra can still be
generated by two generators $a_N$ and $a^\dagger_N$ defined as
\be
a_N=\sum_{n=N}^\infty \sqrt{(n+1)\,\theta}\dm{n}{n+1}\ \ ; \ \
a^\dagger_N=\sum_{n=N}^\infty \sqrt{(n+1)\,\theta}\dm{n+1}{n} \ .
\ee
but now the commutation relation~\eqn{aadcomm} does not hold
anymore, the commutator is not a constant, but it is a
diagonal matrix, and hence it corresponds to a radially symmetric
function.

\section{Fuzzy Chern-Simons and Edge States \label{se:CS}}
\setcounter{equation}{0}

\begin{figure}[htbp]
\epsfxsize=2.5 in \centerline{\epsfxsize=2.5
in\epsffile{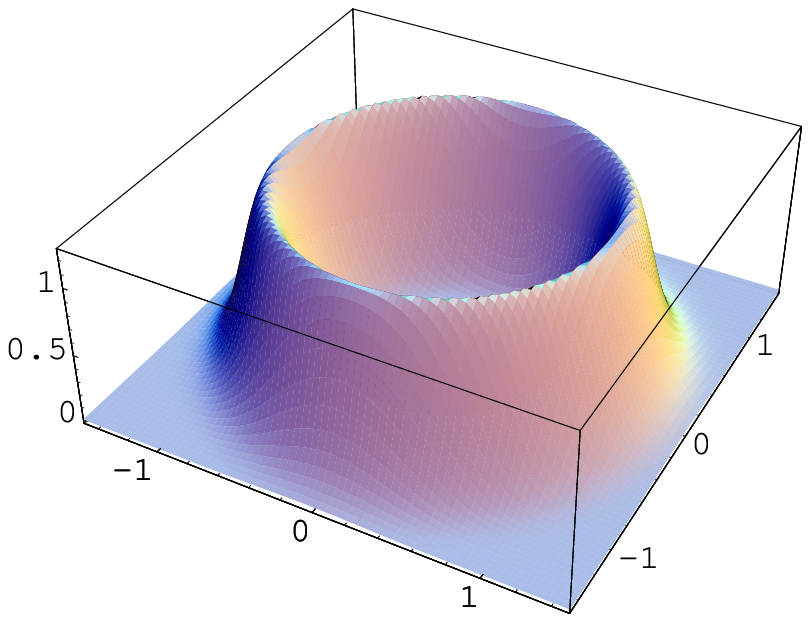} \epsfxsize=2.5
in\epsffile{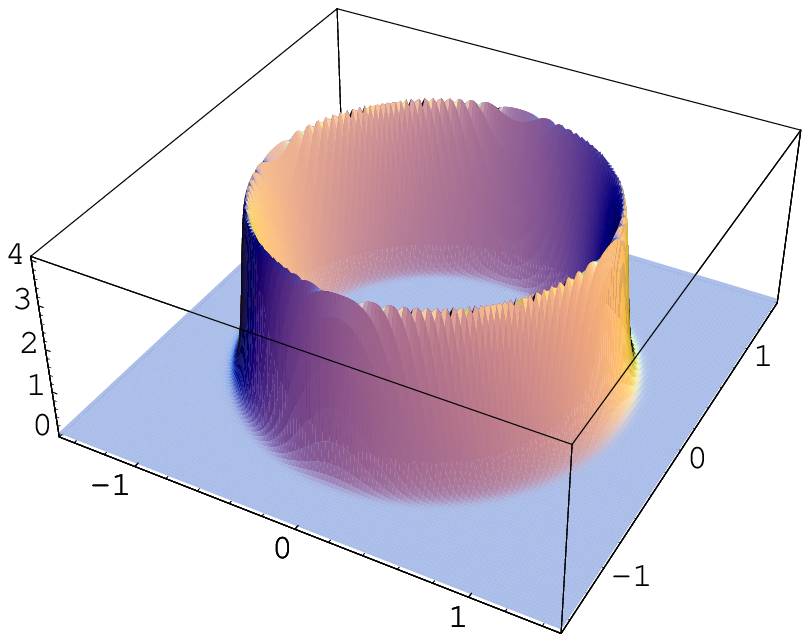}} \caption{\baselineskip=12pt {\it
The edge states $\bra{z}E_N\ket{z}$ for $N=10$ and $N=100$. }}
\bigskip
\label{edge}
\end{figure}

We want to discuss now the possibility to identify in our matrix
approximation on the disc the edge states which play a crucial
role in Chern-Simons theories\cite{edge}. In order to do this we
need a matrix equivalent of the derivatives\cite{fuzzydisc}. We
define
\be
\del_z\Phi=\frac{1}{\theta} [a^\dagger,\Phi]\ \ ; \ \ \del_{\bar
z}\Phi=\frac{1}{\theta}[a,\Phi]  \ .
\ee
Note that in the above expression $a$ and $a^\dagger$ are still
infinite matrices, and therefore if $\Phi$ is an $N\times N$
matrix, $\del\Phi$ is of rank $N+1\times N+1$, this is crucial for
the identification of the matrix algebra with a disc, and means
that the derivatives live in an extended algebra, the extension is
the range of the projector
\be
E_{N+1}=\ket{N+1}\bra{N+1} \ ,
\ee
and heuristically means that if functions vanish on the boundary
of a disc, their derivatives do not necessarily do so. They are
\emph{edge states}. In figure~\ref{edge}
edge states for $N=10$ and $N=100$ are shown.

The definition of a matrix model corresponding to a Chern-Simons
theory is straightforward\cite{BalGuptaKurkcuoglu}:
\be
S_{cs}=\int dx_0\Tr \varepsilon_{\mu\nu\sigma}A_\mu\del_\nu
A_\sigma+\frac 23 A_\mu A_\nu A_\sigma
\ee
where $\mu\nu\sigma=0,z,\bar z$, the $A$'s are elements of the
matrix algebra, and time has not been quantized nor fuzzyfied. A
canonical analysis of this model has been performed
in~\cite{BalGuptaKurkcuoglu} to which we refer for details. We
limit ourselves to note that the gauge invariance of the fields:
\be
A_\mu\longrightarrow A_\mu+(\del_\mu\Lambda+i[A_\mu,\Lambda])
\ee
forces the $A_z$ and $A_{\bar z}$ (but not $A_0$) to live on the
range of $P_N^\theta+E_{N+1}$, that is to contain edge states.
In fact these edge states are the only degrees of freedom of the
theory both in the continuum\cite{edge} and the
matrix \cite{BalGuptaKurkcuoglu,PinzulStern} cases.

\section*{Acknowledgments}
We thank A.P.~Balachandran, B.~Dolan, K,~Gupta, G.~Immirzi,
S.~K\"urk\c{c}\"{u}o\v{g}lu, G.~Landi, G.~Marmo, X.~Martin,
D.~O'Connor, P.~Presnajder, R.~Szabo, A.~Stern and J.~Varilly for
useful discussions. This work has been supported in part by the
{\sl Progetto di Ricerca di Interesse Nazionale {\em SInteSi}}.



\vspace*{6pt}

\begin{thebibliography}{0}

\bibitem{fuzzydisc} F.~Lizzi, P.~Vitale and A.~Zampini, {\it The Fuzzy
Disc}, JHEP {\bf 0308}, 057 (2003) [hep-th/0306247].


\bibitem{BalGuptaKurkcuoglu} A.P.~Balachandran, K.~Gupta and
S.~K\"urk\c{c}\"{u}o\v{g}lu, {\it Edge Currents in Noncommutative
Chern--Simons Theory from a new Matrix Model}, JHEP {\bf 0309}, 007 (2003)
[hep-th/0306255].


\bibitem{PinzulStern}  A.~Pinzul and  A.~Stern,
{\it Absence of the holographic principle in noncommutative
Chern-Simons  theory}, JHEP {\bf 0111}, 023 (2001)
[hep-th/0107179]; {\it A new class of two-dimensional
noncommutative spaces}, JHEP {\bf 0203} (2002) 039
[arXiv:hep-th/0112220]; {\it W-infinity algebras from
noncommutative Chern-Simons theory}, Mod.\ Phys.\ Lett.\ A {\bf
18} (2003) 1215 [arXiv:hep-th/0206095]; {\it Edge States from
Defects on the Noncommutative Plane}, These Proceedings,
[hep-th/0307234].


\bibitem{Madorefuzzysphere} J. Madore,  {\it The fuzzy sphere},
Class. Quant. Grav. {\bf 9} (1992) 69.

\bibitem{GrosseStrohamer}
H. Grosse and  A. Strohmaier, {\it Towards a nonperturbative covariant
regularization in 4-d quantum field theory}, Lett. Math. Phys.
{\bf 48} (1999) 163 [ hep-th/9902138].

\bibitem{Ramgoolam}
S. Ramgoolam, {\it Higher dimensional geometries related to fuzzy
odd dimensional spheres}, JHEP {\bf 0210} (2002) 064 [
hep-th/0207111];
{\it On spherical harmonics for fuzzy spheres in diverse
dimensions,} Nucl.\ Phys.\ B {\bf 610}, 461 (2001)
[hep-th/0105006].

\bibitem{thefuzz} A.~P.~Balachandran and G.~Immirzi,
{\it Fuzzy Nambu-Goldstone physics,} [hep-th/0212133];
A.~P.~Balachandran, B.~P.~Dolan, J.~H.~Lee, X.~Martin and
D.~O'Connor, {\it Fuzzy complex projective spaces and their
star-products,} J.\ Geom.\ Phys.\  {\bf 43} (2002) 184
[hep-th/0107099];
A.~P.~Balachandran, X.~Martin and D.~O'Connor, {\it Fuzzy actions
and their continuum limits,} Int.\ J.\ Mod.\ Phys.\ A {\bf 16},
2577 (2001) [hep-th/0007030];
J.~Medina and D.~O'Connor, {\it Scalar field theory on fuzzy
$S^4$,} hep-th/0212170;
G.~Alexanian, A.~P.~Balachandran, G.~Immirzi and B.~Ydri, {\it
Fuzzy CP(2),} J.\ Geom.\ Phys.\  {\bf 42} (2002) 28
[hep-th/0103023].

\bibitem{Voros} H. Gr\"onewold, {\it On principles of quantum mechanics}, Physica {12}
(1946) 405; J. E. Moyal,  Quantum mechanics as a statistical
theory, Proc.Cambridge Phil.Soc. {\bf 45} (1949) 99; A. Voros,
{\it Wentzel-Kramers-Brillouin method in the Bargmann
 representation}, Phys. Rev. {\bf A 40} (1989) 6814.

\bibitem{edge} E. Witten, {\it Quantum field theory and the Jones polynomial}, Comm. Math.
Phys. {\bf 121} (1989) 351;
G. W. Moore and N. Seiberg, {\it Taming the conformal zoo}, Phys.
Lett. {\bf B 220} (1989) 422;
S. Elitzur, G. W. Moore, A.Schwimmer and N.Seiberg, {\it Remarks
on the canonical quantization of the Chern Simons Witten theory}
Nucl. Phys. {\bf B 326} (1989) 108;
A. P. Balachandran, G. Bimonte, K. S. Gupta and A. Stern, {\it The
Chern Simons source as a conformal family and its vertex
operators} Int. J. Mod. Phys. {\bf A 7} (1992) 5855 [
hep-th/9201048].
\end{thebibliography}
\end{document}